*Title*: Who will import hydrogen in 2050? Global assessment with China and US case studies

*Authors*: Veronika Brooks[a, b,*], Joshua Fragoso García[c], Lin Zheng[c], Viktor Paul Müller[c], Christoph Nolden[a], Dominik Möst [b], Martin Wietschel[c]

*Affiliations*:

[a]Fraunhofer Research Institution for Energy Infrastructures and Geotechnologies IEG, Cottbus 03046, Germany

[b]Chair of Energy Economics, Technical University Dresden, Dresden 01069, Germany

[c]Fraunhofer Institute for Systems and Innovation Research ISI, Karlsruhe 76139, Germany

*Lead author, correspondence: veronika.brooks@ieg.fraunhofer.de

*Highlights*: (3-4)

- Combined meta-analysis and case study modelling approach
- Global hydrogen import potential range from 576 to 1,514 TWh, substantially lower than previous literature.
- Hydrogen trade in 2050 is expected to remain largely regional, with 16 countries emerging as net importers.
- The United States is likely to meet its hydrogen demand domestically, while China will depend on imports by 2050.

*Abstract*

This study assesses the global hydrogen import potential in 2050 by looking at the renewable hydrogen production potential in prospective import-oriented countries. Renewable energy potentials calculated with a GIS based model and 2050 primary energy consumption projections are used to identify candidate importers by comparing it with expected demand. Two approaches are applied: (1) a meta-analysis of literature on hydrogen production potential and demand for the identified countries, and (2) detailed regional analyses for the United States and China. The results suggest limited prospects for a fully global hydrogen market, although certain countries, such as Germany, Italy, and the Netherlands, are likely to remain net importers. By contrast, many Asian countries have enough renewable resources to decarbonise their energy systems. China and the United States may follow divergent pathways: while the United States is well placed to meet its overall demand, using its own resources, it can still benefit from regional connections with neighbouring countries. China is facing strong demand in its eastern regions, so it may source cost-effective renewable hydrogen from foreign countries instead of relying solely on domestic production. Overall, the findings suggest that hydrogen trade will likely remain regionally concentrated, shaped by the interplay of renewable resource distribution, energy demand, other available hydrogen production pathways, and infrastructure constraints.

# 1 Introduction

In recent years, green or sustainable hydrogen has gained prominence as a low-carbon energy carrier. The IPCC highlights its potential to connect sectors and decarbonise hard-to-abate industries[1]. Its role is expanding beyond chemical feedstocks to include long-term energy storage, aviation and maritime transport, and high-temperature industrial processes[2]. Forecasts suggest the global hydrogen market could reach USD 1.4–1.6 trillion by 2050[3], with demand projected between 4 and 15 PWh corresponding to equivalent of 4-11% of global energy use [2,4]. Many countries have published hydrogen strategies that outline domestic demand and, in some cases, export ambitions[5]. These evolving roles are shaped by geopolitical factors, the availability of renewable resources, and the economic context. Yet, future hydrogen trade patterns remain uncertain. Potential hydrogen trade will mainly be influenced mainly by factors such as production costs, the types of products transported, and synthesis or conversion process steps. Additionally, producing low-carbon hydrogen domestically could be more economically attractive than importing it from countries with vast resources, given the losses incurred in the complex supply chain. Monitoring of announced projects on hydrogen supply revealed slower progress compared to a year ago[6]. Despite these delays, recent studies suggest that hydrogen trade could emerge globally, prompting many countries to outline prospective supply chains [7–10].

## 1.1 Hydrogen export-oriented countries

The Hydrogen Council and McKinsey identify key future hydrogen exporters by 2050, including the United States, Canada, North African nations, Australia, the Middle East, and Latin America[8]. They estimate that over half of global hydrogen demand could be met via long-distance transport of hydrogen or its derivatives. While some regions benefit from abundant low-cost renewables, countries like the U.S. and Canada also rely on policy support, such as the US Inflation Reduction Act and Investment Tax Credit [8].

The IEA identifies potential hydrogen trade based on projects that have reached the final investment decision stage, including production, infrastructure, and potential offtakers. According to its *Global Hydrogen Review 2023*, the countries likely to be exporters by 2050 include Australia, the US, Latin America, Africa, the Middle East, and parts of Europe. These projections are based on announced export-oriented projects, although offtakers have been identified for less than a third of the planned capacity[6,10].

Kim et al. 2021 highlight Australia, Saudi Arabia, Argentina, and Russia as potential hydrogen exporters based on their renewable energy potential[7]. Global hydrogen atlases such as HYPAT, PtX Atlas, and EWI identify many more countries that could become hydrogen exporters [11–13]. For example, Wietschel et al. 2024 (HYPAT) presents a global supply curve where 48 countries are selected to export hydrogen to Germany via various transport modes [11].

## 1.2 Hydrogen import-oriented countries

Several studies identify import-oriented regions, most notably countries such as Germany, the Netherlands, and Italy, as well as Japan, South Korea, and Singapore. These countries are characterised by limited domestic renewable energy resources and high industrial energy demand, making them likely candidates for hydrogen imports[7–10,14–16]. Kim et al. (2021) claim that, apart from extensive hydrogen demand, overall reliance on energy imports, such as natural gas, coal, crude oil and its products, and electricity, as well as other factors including high electricity prices and land shortages, plays a role in favouring supply from other countries. One example is a potential future supply from Indonesia to South Korea[7]. The study by Panchenko et al. encompasses a literature review on the global prospects of renewable hydrogen production, additionally defining China, South Africa, the USA, and Chile as hydrogen importers from Australia [15]. Hydrogen Council and McKinsey suggest that China, the United States and India may meet most of their demand domestically but could still import from lower-cost producers such as the Middle East or Australia[8].

In the Global Hydrogen Review 2023 and 2024, the IEA identifies the hydrogen-importing regions based on globally announced projects by matching the suppliers and offtakers. A significant share of projected hydrogen flows is assigned to undefined offtakers or projects without a final investment decision (FID) [9,10]. Export-oriented countries are clearly defined compared to the importing regions. This imbalance suggests an uncertainty in the global hydrogen trade, as there is a gap between supply readiness and a lack of clear demand centres.

Galimova et al. developed a global trade model to identify exporter and importer countries for synthetic fuels, such as e-methane, e-ammonia, and power-to-liquid (PtL), based on levelised costs. The study also assessed market value, trade volume, and geopolitical risks, with a focus on diversifying import portfolios. By 2050, exporters include regions in North and South America, sub-Saharan Africa, China, and Australia; importers are found in Europe, South and Southeast Asia, and parts of Africa. The model assumes fixed demand in 92 countries and does not account for domestic production potential in import-oriented countries. It also restricts each country to either an exporter or importer role, which oversimplifies the real-world dynamics, particularly for large countries like the U.S. or China, where costs and demand vary regionally[17].

## 1.3 Literature gap and Research questions

Many studies identify key hydrogen-exporting countries, but the role of importers remains uncertain [6,10,13,18,19]. While nations like Germany, the Netherlands, Japan, and South Korea are expected to become major consumers, the broader global hydrogen import landscape is unclear [9,10,15,17,20–24]. This study conducts a meta-analysis of national-level literature to estimate long-term renewable hydrogen import potential, focusing on hydrogen as the sole energy carrier. Countries are selected based on renewable resource availability and domestic energy priorities. The analysis includes 2050 projections of hydrogen production and demand. Additionally, two case studies on the USA and China are selected. A detailed analysis on the regional level is therefore conducted to assess their potential as importers, exporters, or both, which remains unclear in the current literature. This study answers the following research questions:

- Which countries could become the main importers of renewable hydrogen worldwide by 2050?
- What will be the size of the international hydrogen trade by 2050, given the hydrogen demand and potential of the identified countries?
- What are the energy policy implications of such developments?

The paper is structured as follows: The methodology starts with the identification and listing of the global import-oriented countries, followed by demand and renewable hydrogen production potential estimation, as well as a meta-analysis and in-depth analysis of two cases, US and China. The results and discussion share the same structure and are divided into three identical parts: meta-analysis, in-depth analysis of case studies and global import potential.

## 2  Methodology

This section outlines the methodological steps used in this paper. First, we describe the identification of countries that may be importing hydrogen in 2050 based on their projected energy deficits (Section 2.1). Second, we present the methodology used to conduct the meta-analysis (Section 2.2). Finally, we elaborate on the methodology applied for the case study calculations of China and the United States (Section 2.3). The overall methodology is shown in Figure 1.

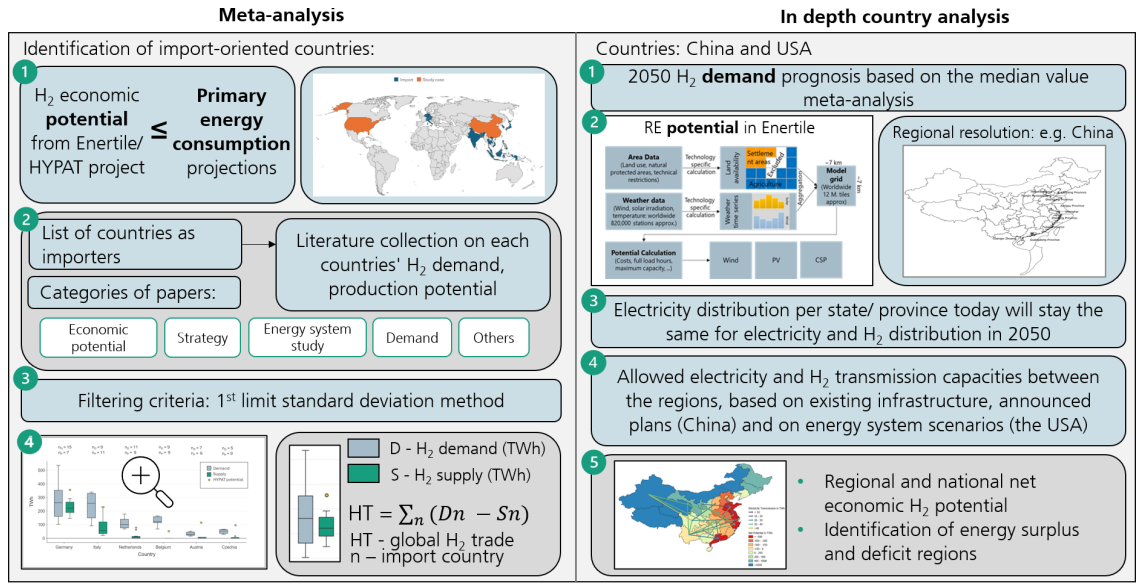

Figure 1: Overall methodology followed in this paper. Left side methodology for the meta-study. Right side, methodology for the in-depth country analysis.

### 2.1  Country identification

The first step in our study is to identify countries with future hydrogen import needs, as shown in Figure 1. We focus on countries that may face an energy deficit when comparing projected primary energy consumption with their renewable electricity generation potential in 2050. Our analysis prioritises electricity because most processes are expected to be electrified to achieve decarbonisation[25]. Electricity is crucial for producing electrolytic hydrogen, the primary focus of this paper. Therefore, in this section, we first describe the methodology used to determine the renewable electricity generation potential, followed by the calculation of energy demand projections. Finally, we outline the process for selecting the import countries (see step 2 on the left side of Figure 1).

### 2.1.1 Renewable potential calculation

Renewable electricity generation potential is assessed for utility-scale PV, concentrated solar power, wind onshore, and wind offshore technologies using the Renewable Potential Calculator 2.0 of the Enertile model on a global scale[26]. The model divides the earth´s surface into approximately 12 million tiles, each 6.5 x 6.5 km². Weather data are assigned to each tile to calculate hourly generation time series for each technology, using ECMWF ERA5 reanalysis data for 2010[27]. To determine the potential installable capacity, land types are assigned to each tile based on the GlobCover 2009 dataset [28]. Rooftop PV was not included in the calculation, as the electricity produced by this technology is difficult to allocate for large scale hydrogen production. Subsequently, land-use factors are then applied to each land type, indicating the proportion of land that can be used for each technology. The model excludes protected areas classified as $I_a$, $I_b$, and II by the International Union for Conservation of Nature and Natural Resources[29]. Results from individual tiles are aggregated as a weighted average and systematically categorised into steps according to their levelized cost of electricity (LCOE), with step 0 representing the lowest cost. 2018 is used as base year for LCOE calculation. Further details of these calculations can be found in the literature [26,30,31].

### 2.1.2 Global demand estimation

The estimation of global demand was taken from Breitschopf et al. (2022). Building on the JRC Global Energy and Climate Outlook 2020, this study provides projections of future energy demand for 172 countries worldwide. For countries not covered by the JRC outlook, the authors used a disaggregation methodology to derive country-level estimates, ensuring consistent and comprehensive global coverage[32].

### 2.1.3 $H_2$ import country identification

To identify countries that could import hydrogen by 2050, we subtracted the calculated energy demand (as detailed in Section 2.1.2) from the renewable energy potential (outlined in Section 2.1.1). Locations with unrealistically high generation costs were excluded by applying a cost threshold of 100 €/MWh to the renewable electricity generation potential, according to the LCOE values calculated with the Enertile model (see Section 2.1.1). This threshold allows for the inclusion of offshore wind potential, which, despite higher capital costs, offers advantages such as fewer acceptance issues and additional generation potential. The analysis assumes that countries prioritise domestic renewable resources and that the global energy system follows a 1.5–2.0 °C decarbonisation pathway by 2050, in line with the Paris Agreement. Table 1 lists the countries projected to face an energy deficit by 2050, where renewable energy potential is insufficient to meet national, demand implying a need for national imports. By excluding countries with deficits smaller than 30 TWh, we identified the regions with the most significant energy deficits: Central Europe, Eastern Asia and Southeast Asia. These areas will be referred to as the European and Asian regions thereafter. Consequently, the analysis focuses on these two regions. Trinidad and Tobago and Bahrain were excluded because they fall outside these categories.

Table 1: Country selection after comparing renewable electricity generation potential with projected demand in 2050. Reported are the total renewable electricity potential (TWh), projected electricity demand, the calculated balance (potential − demand, with negative values indicating a deficit), and whether the country is included in the study (deficit larger than 30 TWh)

| COUNTRY | SUM OF GENERATION POTENTIAL (TWH) | PRIMARY ENERGY DEMAND (TWH) | DEFICIT | CONSIDERED IN THIS STUDY (DEFICIT LARGER THAN 30 TWH) |
|---|---|---|---|---|
| SLOVENIA | 41 | 42 | 1 | ✗ |
| MALTA | 1 | 5 | 4 | ✗ |
| BHUTAN | 16 | 23 | 7 | ✗ |
| BANGLADESH | 431 | 448 | 17 | ✗ |
| EQUATORIAL GUINEA | 35 | 52 | 17 | ✗ |
| BRUNEI | 3 | 21 | 18 | ✗ |
| LUXEMBOURG | 12 | 44 | 32 | ✓ |
| CZECHIA | 205 | 242 | 37 | ✓ |
| NETHERLANDS | 473 | 517 | 44 | ✓ |
| PHILIPPINES | 492 | 541 | 49 | ✓ |
| SWITZERLAND | 92 | 183 | 91 | ✓ |
| BAHRAIN | 12 | 108 | 96 | ✗ |
| TRINIDAD AND TOBAGO | 20 | 129 | 109 | ✗ |
| AUSTRIA | 144 | 259 | 115 | ✓ |
| SINGAPORE | 0 | 248 | 248 | ✓ |
| BELGIUM | 122 | 395 | 273 | ✓ |
| VIETNAM | 820 | 1196 | 376 | ✓ |
| ITALY | 536 | 937 | 401 | ✓ |
| GERMANY | 1488 | 1952 | 464 | ✓ |
| THAILAND | 971 | 1593 | 622 | ✓ |
| MALAYSIA | 371 | 1049 | 678 | ✓ |
| INDIA | 13899 | 14774 | 875 | ✗ |
| SOUTH KOREA | 324 | 2262 | 1938 | ✓ |
| JAPAN | 594 | 2928 | 2334 | ✓ |
| INDONESIA | 2249 | 4615 | 2366 | ✓ |

In addition to the 16 selected countries listed in Table 1, we conducted a more detailed analysis of China and the United States. These two countries were chosen because, although both have sufficient electricity generation potential to meet their demand by 2050, the locations of these resources do not necessarily align with the main demand centres. This raises the question of whether it might be more economically feasible to source energy from neighbouring countries rather than relying solely on domestic production. Furthermore, China and the United States are the world´s second- and third-largest countries by area, the first and second largest economies and the two biggest $CO_2$ emitters[33–35]. Therefore, we considered China and the United States as case studies. Although India is identified as an importer in our preliminary study, given its size and the potential, the possible spatial mismatch between supply and demand, it is a similar case to the United States and China. However, it was excluded from the present analysis due to resource constraints. Future work could expand the scope to include India alongside China and the United States. Figure 2 illustrates the countries selected for this study.

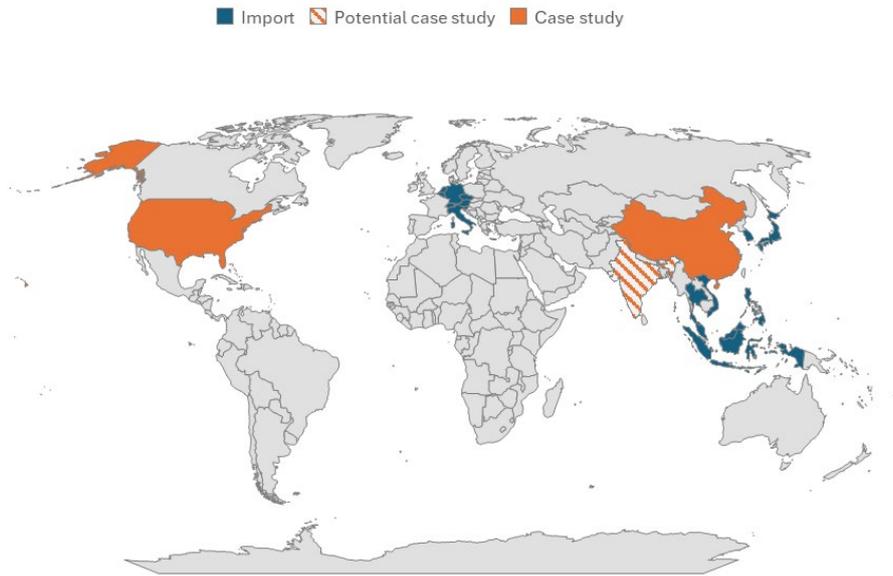

Figure 2: Geographical location of the countries selected for this study. Countries shown in blue indicate potential hydrogen importers, primarily located in two regions. The United States and China are marked in orange and are included as case studies. India is identified as case study but excluded from the paper scope.

## 2.2 Meta-analysis

The import needs of the selected 16 countries (see Step 1 in Figure 1; full list in Table 1) were evaluated through a meta-analysis based on two categories: hydrogen demand projections and hydrogen supply potential in 2050. As shown on the left side of Figure 1, Step 2, the meta-study combines various study types (see Tables 2 and 3), including studies on renewable hydrogen production potential, energy system analyses, and national strategies and roadmaps. We also added our own calculations for green hydrogen production potential based on the HYPAT project[11,31]. The literature was collected according to the following criteria:

- Demand studies: focus on year 2050; the collected studies considered the final hydrogen demand. The discrepancies in the assumptions for the hydrogen sectoral demand led to high uncertainty in the future projections.
- Supply studies: focus on year 2050; renewable hydrogen production potential, energy system studies with 100% renewable electricity, and economic hydrogen production potential. In the energy system optimisation studies, the production potential was assessed as competitive and therefore sufficient to meet the modelled demand.

In total, 388 values were collected and are available in the supplementary documents. All studies from the meta-analysis were categorised into five types: economic potential, energy system studies, strategy, demand, and others to get a better insight into the meta-analysis results. Tables 2 and 3 provide the definitions behind the types of values collected for the meta-analysis. For the third step, we applied the standard deviation method using the first limit to filter out outliers and to narrow the range of the potential import volumes for each country (see Figure 1). In the final step, differences between the quartiles of hydrogen demand and production potential were calculated and presented as ranges, forming the basis for estimating the total global hydrogen import potential.

Table 2: Type of studies evaluated for the meta study. Table includes a definition of each category.

| Type of study | Definition |
| --- | --- |
| Economic potential | The renewable hydrogen potential under 2-3 €/kg (70-100 €/MWh) threshold |
| Energy system | Endogenous domestic renewable hydrogen supply as a result of the optimisation model |
| Strategy | National strategies, targets, and roadmaps related to renewable hydrogen production |
| Demand | National hydrogen demand |
| Others | Other types of studies, where the methodology of obtaining the hydrogen production potential is not available |

Table 3: Overview of the number of values collected per region, country and study type.

| World Region | Country | Type value | | | | | |
| --- | --- | --- | --- | --- | --- | --- | --- |
| | | economic potential | system analysis | strategy | demand | other | Total |
| Europe | Germany | 4 | 8 | 0 | 20 | 0 | 32 |
| | Netherlands | 1 | 9 | 0 | 18 | 0 | 28 |
| | Italy | 6 | 8 | 0 | 13 | 0 | 27 |
| | Belgium | 4 | 6 | 0 | 13 | 0 | 25 |

|  | Austria | 1 | 6 | 0 | 9 | 0 | 16 |
|---|---|---|---|---|---|---|---|
|  | Czechia | 4 | 6 | 0 | 7 | 0 | 17 |
| Asia | Japan | 5 | 14 | 1 | 35 | 0 | 55 |
|  | South Korea | 5 | 1 | 2 | 15 | 0 | 23 |
|  | Indonesia | 4 | 8 | 1 | 18 | 1 | 32 |
|  | Thailand | 1 | 3 | 0 | 13 | 2 | 19 |
|  | Philippines | 1 | 1 | 0 | 6 | 1 | 9 |
|  | Vietnam | 2 | 6 | 2 | 8 | 1 | 19 |
|  | Singapore | 1 | 1 | 0 | 5 | 1 | 8 |

## 2.3   In-depth analysis with case studies

Following the methodology description of the meta-study, this section presents the methodology for the in-depth analysis of the two case studies, China and the United States.

The electricity and hydrogen demand for 2050 was calculated using the meta-study described in Section 2.2. The median values from this study were then distributed among US states and Chinese provinces based on their current electricity consumption shares[36,37]. It is therefore assumed that these regional shares will remain constant by 2050. Assuming a 70% efficiency, hydrogen demand was converted into electricity equivalents, enabling both energy carriers to be evaluated consistently.

The renewable energy potentials were aggregated at the province and state levels, as illustrated in Step 2 on the right-hand side of Figure 1. The values correspond to the same potentials described in Section 2.1.1, re-aggregated to these administrative units. For offshore wind, a share of the exclusive economic zone was assigned to each coastal state or region. The potentials were capped at 100 €/MWh. A map showing the allocation for both the United States and China is provided in Appendix X. As a final step (Step 4, right-hand side of Figure 1), we consider electricity (USA) and electricity and hydrogen (China) transmission between the states and regions. For the United States, we consider the transmission capacities simulated in the "National Transmission Planning Study" from NREL. The scenario AC_DemMd_90by2035EP__core was selected[38,39]. This scenario is a conservative one that permits new interregional connections but no DC transmission lines. For the case of China, we considered both planned high-voltage DC lines and hydrogen pipelines from different sources[40–44].

# 3     Results

This section is divided into three parts. The first part presents the results of the meta-analysis for the European and Asian regions (Section 3.1), followed by the second part, which includes two case studies: China and the United States (Section 3.2). The third subsection (Section 3.3) shows our results in the context of the global hydrogen market, by deriving the quantities for international trade.

## 3.1     Regional meta-analysis

### 3.1.1     Hydrogen-importing countries of the European region

The meta-analysis of the European region focuses on Germany, Italy, the Netherlands, Belgium, Austria and the Czech Republic. They all exhibit the largest renewable energy deficits, mainly due to high primary energy consumption and insufficient land availability, resulting in limited low-cost renewable potentials. Figure 3 summarises the results of the meta-analysis on the hydrogen demand projections and the renewable hydrogen production potential (see supplementary documents for a detailed list). Subsequently, the number of values for each category (supply or potential) for each country should be at least five to analyse the box plots. Countries that do not meet this criterion are excluded from the analysis. Figures 3 and 4 include the number of values considered for each box.

Among the countries analysed, the future projections in Germany and Italy show the highest demand and supply potentials with the largest spread, especially the hydrogen demand projections. This underlines the considerable uncertainty of hydrogen demand in these two countries. In Germany, projected hydrogen demand spans from 100 to 550 TWh, while the domestic renewable hydrogen production potential ranges between 150 and 300 TWh. For Italy, hydrogen demand is estimated to range from 100 to 350 TWh, whereas the production potential remains more constrained, predominantly between 20 and 129 TWh, with potential to reach up to 230 TWh in certain scenarios. The production potential for the Netherlands, Belgium, Austria and Czech Republic does not exceed 15-20 TWh, due to its limited economic RE availability and the dominance of studies with system analysis.

Additionally, the orange markers in Figures 3 and 4 represent the economic potential for hydrogen production obtained from the HYPAT project. The economic threshold of 3 €/kg (100 €/MWh) was used to quantify the amount of hydrogen produced domestically, which aligns with projected hydrogen production costs reported by Bühler et al. 2025 [45]. The levelized cost of hydrogen (LCOH) methodology was developed and described by Mendler et al. [31]. For Austria and the Czech Republic, based on the HYPAT projections, the demand should be able to be covered domestically. However, the meta-analysis, based on the literature with a high share of energy system studies, indicates importing hydrogen is more economical to cover the demand.

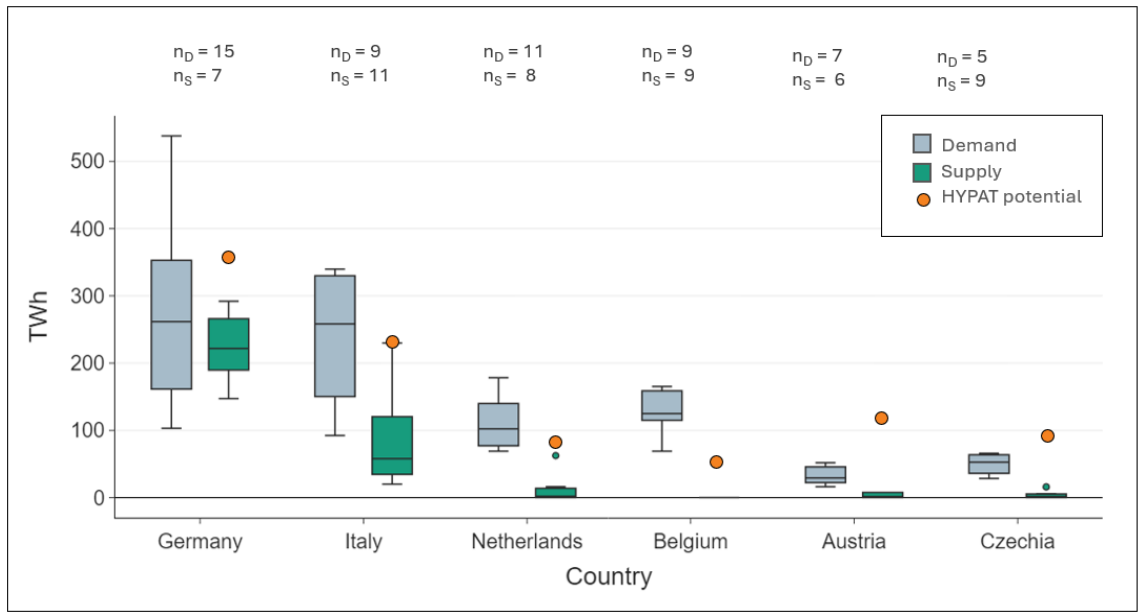

Figure 3: Literature overview of hydrogen supply potential and demand in selected European countries in 2050, with filtering based on the first standard deviation limit. $n_D$ and $n_S$ denote the number of sample values for demand and supply potential, respectively, after filtering[13,31,46–60]. (Full list of literature attached in the supplementary documents)

### 3.1.2 Hydrogen-importing countries of the Asian Region

The meta-analysis of the Asian region shows that the import potential is less distinct than the clearly defined patterns observed in Europe. Countries such as Japan, South Korea, Thailand, Indonesia and Vietnam, see Figure 4, were selected for their significant deficit between renewable energy potential and primary energy consumption. The results do not necessarily show the high hydrogen import potential as some studies concluded in the literature review section. After applying the first standard deviation limit, only Thailand showed slightly higher hydrogen demand than renewable hydrogen production potential. Based on the detailed literature review, countries such as Japan, South Korea, and Indonesia could theoretically decarbonise their hydrogen demand through domestic renewable energy resources, as well as their high reliance on nuclear energy.

Nevertheless, the economic feasibility of these resources remains a critical factor. For instance, while Japan possesses significant offshore wind potential, much of it is in remote areas, therefore making it less economically attractive [61,62]. Countries such as Singapore and the Philippines are excluded from the analysis due to an insufficient number of data points (fewer than five).

In addition, in the Asian region it was found that the hydrogen strategies or roadmaps adopted at government level, e.g. in Japan, South Korea and Vietnam, are several times higher than the values collected in the literature, which are in many cases the expected demand in a cost-optimal energy system (see Figure 3, grey dots represent values from national hydrogen strategies and roadmaps). This indicates the ambitions of these countries to establish a national hydrogen economy or play a significant role in the international hydrogen trade. If such values are taken as a signal to international hydrogen suppliers, then import requirements will increase strongly. Additionally, demand in some countries may be underestimated due to limited diversification of data sources, with most estimates derived from JRC projections [63]. These projections report hydrogen demand as energetic hydrogen, excluding its use as a feedstock. Consequently, the actual demand in these countries could be higher, though it would not reach the levels outlined in the $H_2$ strategies.

The hydrogen production potential from the HYPAT project (orange markers in Figure 4) represents the economic potential for hydrogen production bellow 3 €/kg (100 €/MWh). In Japan, the renewable energy potential could reach 800 TWh, which could cover the demand by unlocking significant offshore wind resources. Due to the great depth of the sea, floating turbines could be an option, although this would raise the cost significantly [62]. For Thailand and Indonesia, the economic hydrogen production potential is estimated at 335 TWh and 835 TWh, respectively, which could, according to selected studies, cover the demand.

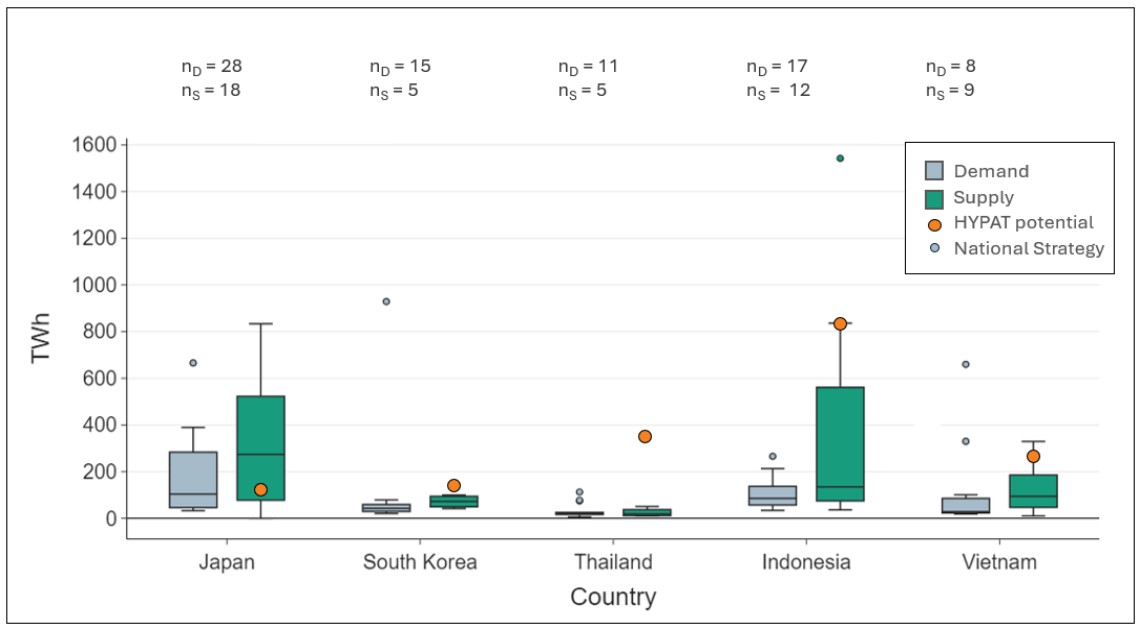

Figure 4: Literature overview of hydrogen demand and supply potential in selected Asian countries in 2050, with filtering based on the first standard deviation limit. $n_D$ and $n_S$ denote the number of sample values for demand and supply potential, respectively, after filtering [20,24,64–83]. (List of literature attached in the supplementary materials)

## 3.2 Focus countries

The literature review includes detailed analyses for the United States and China. Although both countries are expected to account for a substantial share of global hydrogen demand, they also possess significant net production potential and, at first glance, do not appear to be prospective importers (Figure 5). However, due to their large geographical size, it is important to assess renewable energy availability at a higher spatial resolution to identify optimal production locations and major demand centres. The findings from these two case studies are presented below.

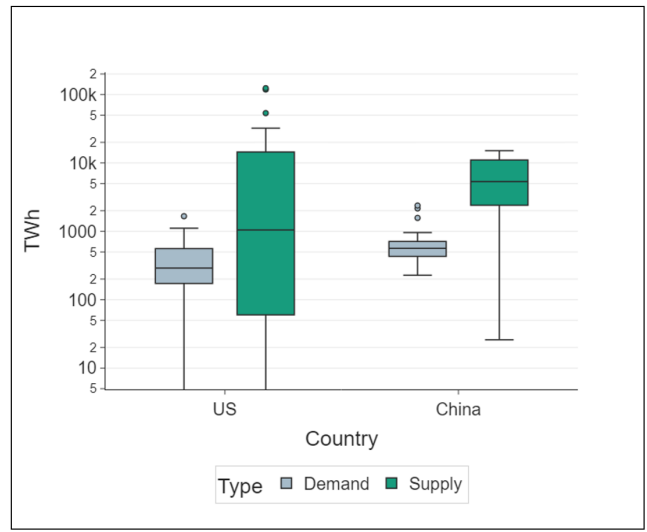

Figure 5: Literature overview on hydrogen supply potential in comparison to hydrogen demand in the US and China, 2050, including the filtering based on the 1st standard deviation limit. [12–14,31,48,63,65–68,84–92] (List of literature attached in the supplementary materials)

### 3.2.1 China case study

Figure 6 illustrates the geographical net potential of renewable energy resources in various Chinese provinces, representing the difference between the total available renewable energy and the total electricity and hydrogen demand in each province. The net potential includes the effect of the planned electricity transmission lines and hydrogen pipelines. Hydrogen is converted into an electrical equivalent. Hydrogen demand, when converted into electricity, accounts for 5.4% of the total demand.

Most of the southeastern provinces, where many demand centres are located, have a significant deficit in renewable energy availability. The regions with the largest deficits include Shandong, Zhejiang, and Guangdong, each with a deficit exceeding 500 TWh. In addition, Hebei, Henan, Anhui, and Fujian each have a deficit of between 300 and 500 TWh. The central part of China has a balance between potential and demand, according to our projections.

Conversely, the northwestern provinces of Xinjiang, Tibet, Qinghai, Gansu, and Inner Mongolia have a high surplus, with each of these regions having a renewable energy potential of over 5,000 TWh. Although the planned transmission and gas infrastructure is designed to transfer energy from these resource-rich western provinces to the deficit-prone eastern regions, the overall shortage in the east remains a significant challenge.

.

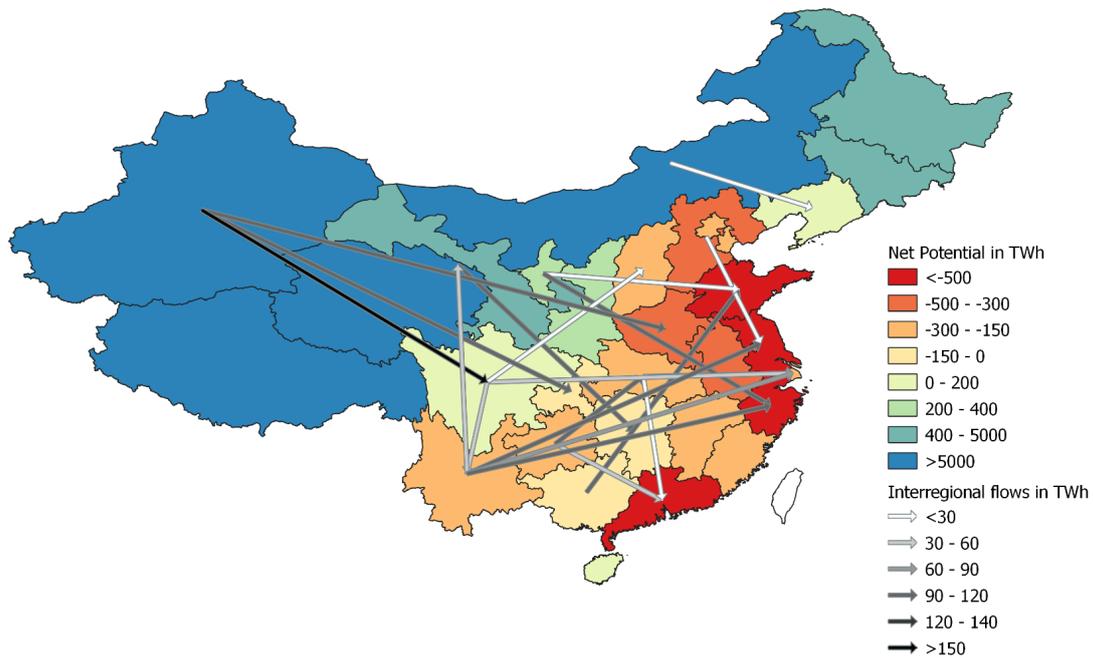

Figure 6: Net potential in Chinese regions (TWh). This accounts for the generation and demand of electricity and hydrogen (converted into electricity equivalents), as well as cross-regional imports and exports. The colour scale shows regional net potential, while the arrows -shown in grayscale- represent the magnitude of the combined electricity and hydrogen flows between regions.

Figure 7 illustrates the potential for renewable electricity generation by levelised cost of electricity (LCOE) tier, alongside provincial electricity demand across China. The electricity demand includes the additional energy required for hydrogen production within each province, further influencing the regional energy balance. Five regions, Xinjiang, Inner Mongolia, Tibet, Qinghai, and Gansu, stand out due to their exceptionally high surplus in renewable generation potential. Each of these provinces has the capability to generate over 3,500 TWh at an LCOE below 20 €/MWh. Their vast renewable resources position them as potential energy hubs, capable of supplying not only local demand but also energy-deficient regions through transmission networks. This is also underlined by the number and the scale of existing and planned transmission lines and pipelines. However, to cover the deficit in energy-deficient regions, more large-scale transmission lines and pipelines will be required.

On the other hand, 14 provinces face a persistent generation deficit, even when considering higher-cost renewable generation options with LCOEs of up to 100 €/MWh. This highlights the challenges of balancing supply and demand across the country, as renewable resources and demand centres are unevenly distributed and the (planned) transmission capacity between them is still far behind what is needed. As previously noted in Figure 6, the provinces with the highest deficits include Shandong, Zhejiang, Guangdong, Hebei, Henan, Anhui, and Fujian. These regions will continue to rely on energy imports despite planned transmission networks, underscoring the need for further infrastructure development and energy diversification strategies. Moreover, the future energy demand in certain coastal provinces may be covered not only by other provinces in China but also by other countries through ships, as is currently the case for coal and LNG.

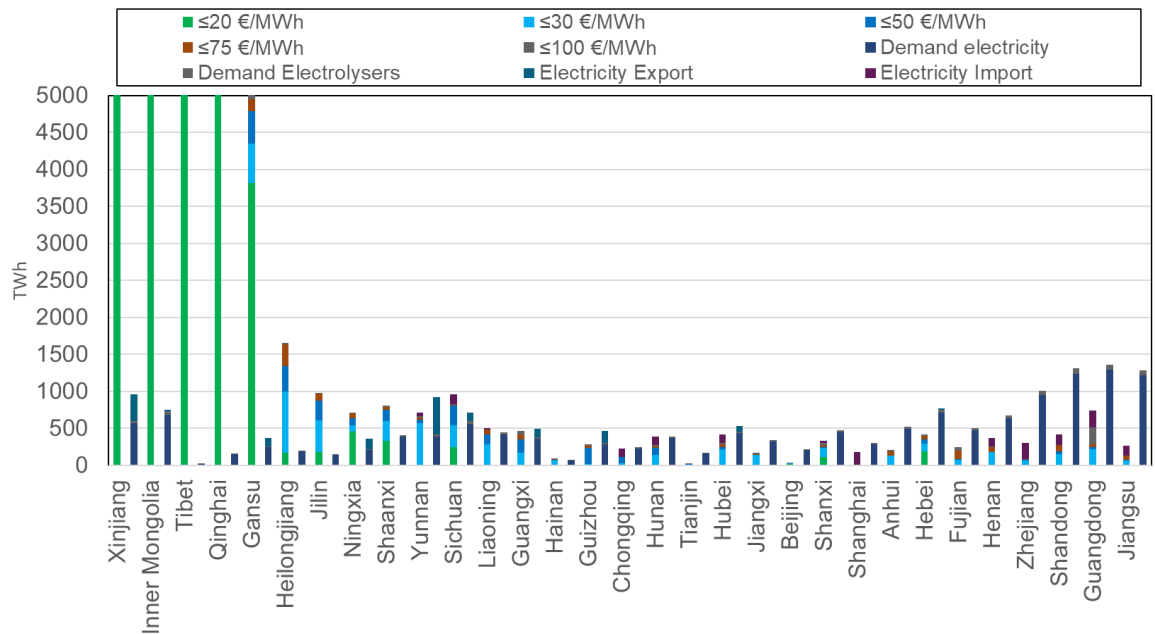

Figure 7: The potential for renewable electricity generation and the projected demand for electricity in China in 2050, broken down by province. Generation potential is shown by the levelised cost of electricity (LCOE) ranges. Demand values include hydrogen demand converted into electricity equivalents. Inter-regional imports and exports are taken into account in the regional net balance.

### 3.2.2 USA case study

Figure 8 illustrates the geographical net potential of renewable energy resources across different U.S. states. Hydrogen demand, when expressed in terms of electricity, accounts for approximately 6.7% of the total electricity demand in 2050. All U.S. states exhibit a net surplus of renewable energy potential; no state faces an energy deficit. However, there are notable regional disparities in the available net potential. States along the East Coast have the lowest potential, ranging between 0 and 200 TWh. In contrast, Texas and New Mexico each have net potentials exceeding 2,000 TWh.

Figure 9 presents the renewable energy generation potential by cost step in comparison to the electricity demand for the United States. Texas stands out with more than 2,500 TWh of potential available at a generation cost of less than 20 €/MWh. Several Midwestern and Western states, including Kansas, Arizona, Wyoming, Colorado, and Nevada, also have substantial potential exceeding 1,000 TWh at the same cost threshold.

On the other end of the spectrum, states such as Connecticut, New Hampshire, Rhode Island, and Vermont exhibit the lowest potential for renewable energy. Nevertheless, these lower values are generally sufficient to meet local demand.

The inclusion of interstate electricity transmission allows for the redistribution of electricity from high-surplus states to those with lower surplus levels, thereby enhancing overall system balance and regional adequacy.

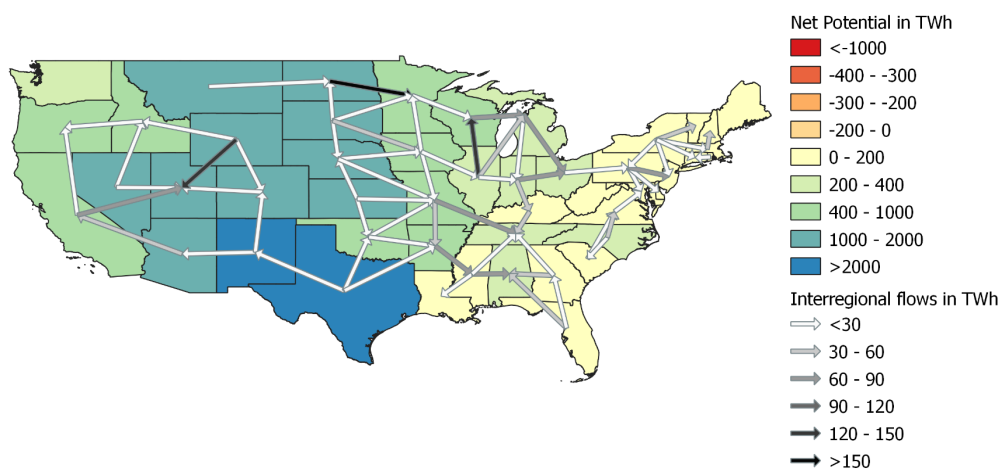

Figure 8: Net potential in the United States per state (TWh). This accounts for the generation and demand of electricity and hydrogen (converted into electricity equivalents), as well as cross-regional imports and exports. The colour scale

shows regional net potential, while the arrows -shown in grayscale- represent the magnitude of the electricity flows between states.

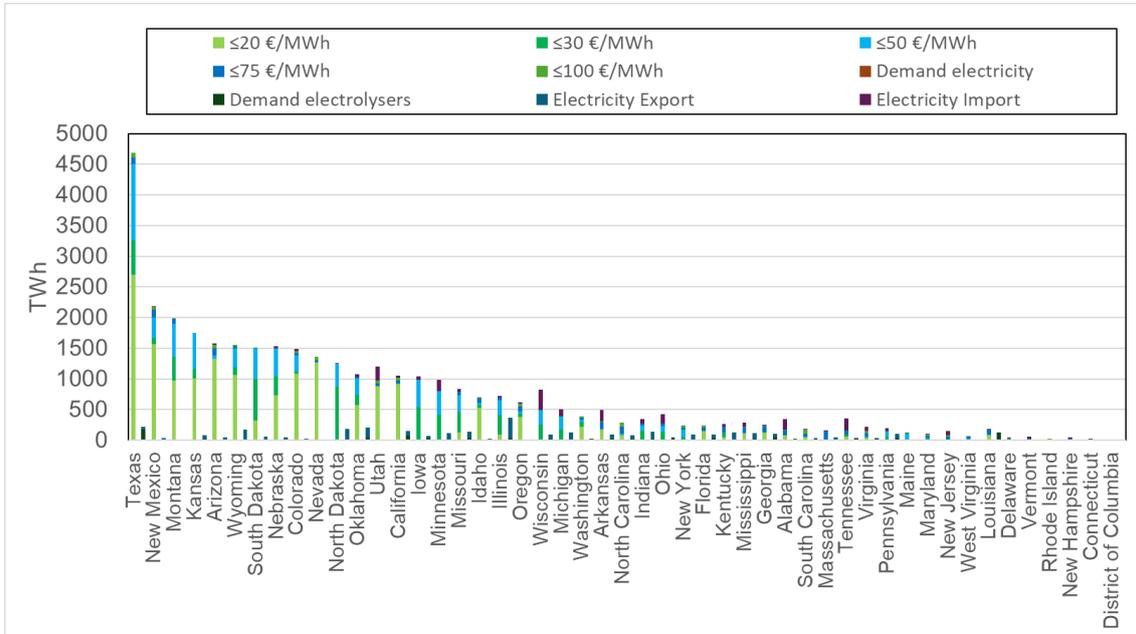

Figure 9: The potential for renewable electricity generation and the projected demand for electricity in the United States in 2050, broken down by state. Generation potential is shown by the levelised cost of electricity (LCOE) ranges. Demand values include hydrogen demand converted into electricity equivalents. Inter-regional imports and exports are taken into account in the regional net balance.

## 3.3  Global hydrogen import potential

The scale of hydrogen imports is calculated based on the demand–supply gap. Table 4 presents various variations of the gap between demand and production potential values from a meta-analysis. As a result, the total import demand spans from 302 to 1,239 TWh. The results show that the European countries are mostly driving the need for global hydrogen share, whereas in the Asian region in only few 2050 projections hydrogen import may arise.

The Chinese provinces identified as deficit regions in Section 3.2.1 have a combined hydrogen demand of around 550 TWh. We assume that at least half of this demand (around 275 TWh) will be met through imports. By contrast, all U.S. states possess sufficient renewable potential to meet their own demand. Consequently, hydrogen imports are expected to be negligible or non-existent.

The long-term outlook for hydrogen could shift with the implementation of targeted policy measures, which essentially concern international commitment to climate protection goals. However, it remains unlikely that the hydrogen economy will scale up to a global level, enabling international trade, as hydrogen transport costs will be significantly higher than those of LNG transportation. In contrast to the oil market, the natural gas market is therefore strongly characterised by regional structures, and this will be even more pronounced in a hydrogen market due to the significantly higher transport costs of hydrogen.

Table 4: Difference between demand (D) and potential (P) TWh (with 1$^{st}$ std. Deviation filter). Positive values refer to the import demand of each country, negative values to the additional hydrogen production potential after covering domestic demand.

| 2050 | Import projections via difference demand and supply TWh | | | |
|---|---|---|---|---|
| Country | Med | Ave | Q3 D - Q1 P | Q1 D - Q3 P |
| Germany | 40 | 63 | 151 | -88 |
| Italy | 200 | 153 | 291 | 48 |
| Netherlands | 100 | 101 | 136 | 69 |
| Belgium | 125 | 130 | 156 | 120 |
| Austria | 29 | 29 | 40 | 17 |
| Czechia | 52 | 47 | 63 | 34 |
| Japan | -118 | -140 | 182 | -427 |
| South Korea | -28 | -25 | 5 | -56 |
| Indonesia | -30 | -238 | 69 | -426 |
| Thailand | 1 | 11 | 39 | -19 |
| Philippines | -15 | -73 | -7 | -111 |
| Vietnam | -55 | -38 | 87 | -136 |
| Singapore | 18 | 16 | 20 | 13 |
| European region | 546 | 523 | 837 | 288 |
| Asian region | 19 | 27 | 402 | 13 |
| Total only import | 565 | 550 | 1239 | 301 |
| Total with case studies | 840 | 825 | 1514 | 576 |

# 4 Discussion

The meta-analysis indicates that international hydrogen trade will grow but remain limited before 2050. In line with our findings, analyses by the Hydrogen Council, McKinsey, IEA, and IRENA suggest that large-scale multinational projects are unlikely before 2030, constrained by weak investment frameworks and uncertainty in energy demand and decarbonisation pathways[8,9,48].

While there is a clear focus on global hydrogen export projects, a significant shortage of committed off-taker countries remains[10]. A similar trend applies to the numerous potential hydrogen-exporting countries, given the vast renewable energy potential available across many regions. This contrasts sharply with conventional energy markets, such as oil and natural gas, where hydrocarbon deposits are geographically concentrated in a few regions. As a result, traditional energy markets exhibited higher market values and intensified competition among consumers. In the emerging hydrogen paradigm, the dynamic may shift toward the consumer side, compelling suppliers to reduce costs and offer additional environmental benefits.

From a national level, country-specific hydrogen strategies indicate a growing demand for renewable-based electrolytic hydrogen by 2030. However, despite the large renewable energy potential, the required global electrolyser capacity remains insufficient, as scaling up production necessitates expanded manufacturing capabilities and requiring up to a decade for feasibility studies and project preparation [93]. Hydrogen production pathways vary, and some countries, under their decarbonisation policies, accept hydrogen produced via steam methane reforming with carbon capture and storage as well as hydrogen from nuclear-based electricity. This could significantly reduce the scale of hydrogen traded on a global level, as countries will prioritise local solutions.

The discussion is also divided into three parts: a meta-analysis of European and Asian import-oriented countries, case studies on China and the USA, and an examination of global hydrogen import potential.

## 4.1 Meta analysis

Many global studies assessing the prospects of the hydrogen market tend to overlook the region-specific hydrogen production potential in import-oriented countries. This paper reviews literature to identify trends shaping major players as hydrogen importers, exporters, or both. The findings are considered in the context of regulatory and policy frameworks, in which import targets for some countries appear to be overestimated, sometimes by severalfold, compared with system studies aiming for the lowest energy system costs for the same countries. This mismatch between policy-driven targets and energy system studies hinders efficient decision-making. While ambitious targets risk unnecessary investment, actual developments in energy transition may also fall short of even the most conservative forecasts.

### 4.1.1 European countries

The meta-analysis of selected import-oriented countries in the EU (Germany, Italy, the Netherlands, Belgium, Austria, and the Czech Republic) revealed a consistent trend: hydrogen demand exceeds the local renewable hydrogen production potential, indicating that they can potentially be strongly import-oriented (See Figure 3). The results are shaped by differences in the availability of literature. For example, Germany has the largest number of collected studies (see Supplementary Materials and Table 3), spanning economic hydrogen production potential, policy and national strategies, and optimisation analyses. For the Czech Republic and Austria, most studies are energy system analyses, which may underestimate the total hydrogen production potential but indicate the economically competitive volumes that can be produced domestically.

Several studies suggest that Europe could achieve energy independence by fully utilising its high renewable potential within a pan-European energy system [13,94–96], which is consistent with our findings. The import-oriented countries are located mainly in the northwestern and central Europe, characterised by high energy demand, particularly in the industrial sector [97]. Other European countries, such as Spain, could supply renewable hydrogen to the deficit countries, complementing results from Lenivova (2022), where the Spanish economic renewable hydrogen potential can be comparable to that of Morocco [95]. Hydrogen imports could also originate from the MENA region [46,96,98], as well as other hydrogen production pathways that could influence the scale of imports required. For example, Norway and the UK are basing their hydrogen strategies on the production of blue hydrogen (produced through steam methane reforming (SMR) with carbon capture and storage (CCS)-based technology) and renewable hydrogen for export to countries such as Germany and the Netherlands. In the near term, blue hydrogen is considered more promising than electrolytic hydrogen, due to its lower costs and earlier deliveries [99].

### 4.1.2 Asian countries

The meta-analysis for the selected Asian countries revealed different results from those for the European countries, where the potential for hydrogen import is less evident (Figure 4). While several studies position Japan and South Korea as $H_2$ importers, our findings suggest that there is sufficient technical renewable energy to meet demand and decarbonise their energy systems. Nevertheless, national strategies in South Korea, Japan, Vietnam and Singapore project substantial import volumes, highlighting a divergence between technical potential and policy ambitions.

Consistent with these projections, several studies examine hydrogen supply chains from Australia to Japan or to South Korea [12,13,100–102]. The PtX Atlas developed by Pfenning et al., estimates economic supply at €120–140 /MWh, depending on the shipping carrier. Kim et al. report an LCOH of €192 /MWh-$H_2$ for the Australia–Korea route, with production and liquefaction accounting for €101 and 36 /MWh-$H_2$, respectively. In Brändle et al. find that due to geographical constraints and limited infrastructure, imports would likely be shipped, adding €59–87 /MWh-$H_2$ in transport costs making domestic hydrogen in Japan (optimistic €51 /MWh-$H_2$ by 2050) potentially competitive.[13,26]. The HYPAT project estimates the economic potential for hydrogen production in Japan to be less than 100 TWh under an 90€/MWh (or 3 €/kg)[11], supporting the possibility of competitive domestic supply. However, avoiding imports entirely would double Japan's power demand by 2050 compared with 2019. This aligns with Burandt (2021), who emphasises that even in countries with poor renewable resources, domestic electrolytic hydrogen can play a balancing role in the energy system [103]. Overall, our results indicate that while imports may support diversification and security of supply, they are unlikely to be the primary driver for achieving net zero in these Asian economies.

South Korea is often identified as a potential net importer of hydrogen, with a national strategy projecting demand in the mobility, power generation, and industrial sectors. Renewable hydrogen production potentials (mainly PV, onshore, and offshore wind) appear sufficient to meet demand by 2050; however, costs are expected to remain high due to poor wind conditions and limited land availability.[104] In recent years, South Korea has pursued international hydrogen cooperation, signing memorandums of understanding (MoUs) with Canada, the USA, and Australia [105]. The country also aims to secure a niche in fuel cell and fuel cell electric vehicles (FCEV) manufacturing and become a global supplier.

In Indonesia, high geothermal and hydro resources can provide stable baseload renewable electricity, complemented by extensive PV potential of ~1,200 TWh [106]. By contrast, wind potential is the lowest among the Philippines, Thailand, and Vietnam. Despite this substantial renewable base, large-scale hydrogen production would require additional storage, including batteries and underground hydrogen storage [106]. Similarly, Thailand, the Philippines, and Vietnam each have significant PV potential of 300–500 TWh, which is sufficient to meet domestic hydrogen demand, while their wind resources remain limited and costly due to low full-load hours (<600 FLH). These patterns highlight how resource profiles, rather than total renewable potential alone, influence the feasibility of hydrogen production across Southeast Asia[106].

## 4.2  Focus Countries

This subsection discusses our results in the context of China and the United States, highlighting their respective needs for hydrogen imports.

### 4.2.1  China case study

In recent years, China has been one of the countries with the most significant expansion of renewable energy (RE) technologies. However, the uneven geographical distribution of RE resources across its territory has created substantial regional imbalances between generation and demand, particularly between the resource-rich northwestern regions and the high-demand eastern and coastal provinces. This discrepancy necessitates the development of large-scale transmission infrastructure to facilitate the transfer of electricity from production centres to consumption centres.

In addition to the RE technologies considered in this study, the Chinese government has announced plans to install up to 500 GW of nuclear power capacity. Assuming an FLH value of 7,500 hours per year, this capacity would generate an annual electricity output of around 3,750 TWh. According to our analysis, the cumulative electricity deficit in the eastern regions is estimated to be around 7,000 TWh, indicating that the planned nuclear capacity alone could potentially supply roughly half of this deficit. Additionally, although the national policy does not set an explicit target for CCS deployment, it is mentioned as one of the key development fields and may potentially play a significant role in China's decarbonisation and possibly also in hydrogen supply.

Furthermore, the hydrogen demand in the deficit regions is estimated to be around 550 TWh. This demand may need to be partially met through hydrogen imports, potentially from regions with abundant resources, such as Australia, where sufficient renewable electricity generation exists to support large-scale hydrogen production and export.

The calculated offshore wind potential in our study amounts to slightly above 900 TWh across all provinces, whereas other studies report values exceeding 12,000 PWh[107]. A key difference is that our assessment limits offshore areas to a water depth of up to 50 m, while Song et al. (2021) consider depths of up to 60 m. This additional 10 m results in a substantially higher estimated potential. However, as noted by Song et al. (2021), areas at a depth of 60 m are located farther from the coast, making them less suitable for local electricity and hydrogen supply. As offshore floating turbines continue to develop in the future, water depth is expected to become a less limiting factor [61]. This progress is expected to substantially increase the available resource potential and cover the hydrogen and electricity in the deficit provinces (see Figure 6). This might reduce the demand for hydrogen imports to China even further.

Our results are consistent with those of Meng et al. (2024), who clearly identify regions with high renewable energy potential[108]. This spatial distinction is delineated by the so-called Hu's hydrogen line. As illustrated in Figure 6, our results similarly highlight the differentiation between regions rich and poor in renewable energy resources. Meng et al. (2024) further suggest that areas with abundant renewable potential could serve as exporters, supplying regions such as Europe.

### 4.2.2 USA

In contrast to China, the United States does not exhibit clearly defined regions with structural energy deficits. However, states along the East Coast utilise nearly the entirety of their available renewable potential up to a cost threshold of 100 €/MWh. Given this limited in-state potential, these regions may increasingly rely on electricity imports from areas with more abundant resources. One such option is strengthening interconnections with Midwestern states, which offer significantly lower-cost renewable energy potential.

Another strategic alternative, though not considered in this paper, is the further expansion and utilisation of existing transmission interconnections with Canada. As of today, the total capacity of these cross-border connections stands at approximately 18 GW. The electricity exchange, both exports and imports, totals 77 TWh. The grid is expected to be expanded by 5 additional GW. The eastern coast of the United States, which is one of the regions with lower available RE potential, is currently connected to the province of Quebec. This permits the export of renewable electricity from hydroelectric plants to the United States. Even if no plans currently exist to expand the interconnection capacity with the United States and Quebec. Quebec plans to expand its hydro generation capacity by 9 GW in 2035 and strengthen its internal interconnection capacity. This may indirectly benefit the East Coast as a larger amount of clean electricity may be available to be transported[109,110]. Similarly, Mexico is closely connected to the United States. Currently accounting for only 0.14% of electricity demand, this connection could be strengthened in the future[111].

In addition to the current and planned electricity interconnections, the region has also begun planning a common framework for hydrogen trade. In 2023, Mexico, the United States, and Canada pledged to explore standards for clean hydrogen. Such standards will facilitate the exchange of clean hydrogen within the region, enabling the creation of a regional hydrogen market. This may be particularly relevant for Canada, as it is closer to regions that may need to rely on high-cost renewable energy sources.

In terms of non-variable baseload capacity, the United States had approximately 94 GW of installed nuclear power as of 2022, with plans underway to substantially expand this figure, potentially tripling or even quadrupling it through the deployment of new plants across the country.[112,113] A recent NREL study highlights the siting potential of both retired coal plants and existing or retired nuclear facilities, assuming the use of 600 MW nuclear units: retired coal sites could accommodate an estimated 174 GW of new capacity, while nuclear sites could host an additional 60 GW[114]. These locations offer the advantage of pre-established community acceptance of power generation infrastructure, thereby reducing potential siting and permitting barriers. Several states, including Pennsylvania, New York, Virginia, and Georgia, were identified as having considerable nuclear potential; while their renewable resources are sufficient to meet future needs, their margins are narrow, making nuclear a valuable complement. Other states, such as Texas and Ohio, could also benefit, with Texas positioned to become a major energy supplier by leveraging both its nuclear siting potential and its vast renewable resources.

## 4.3  Global hydrogen import potential

Based on the results of the meta-analysis, the global potential for importing hydrogen is estimated to be between 576 and 1,514 TWh. This accounts for a maximum of 10% of the projected global demand for hydrogen. This estimate is an order of magnitude lower than the values reported in the literature. For comparison, McKinsey & Company estimates that by 2050, demand for clean hydrogen can reach 12.5 PWh, of which more than half (6.3 PWh) will be transported over long distances. In 2050, less than half of the traded hydrogen will be delivered via long-distance pipelines, mainly concentrated in the North American region, North Africa, Europe, and China, while the rest will be realised in the form of shipments of hydrogen derivatives [8]. Fakhreddine et al. estimates that global hydrogen demand will reach 234 Mt (7.8 PWh) by 2050, with around 30% (2.34 PWh) expected to be traded internationally which 1.6–4 times higher than in our analysis[112].

Despite the potential for renewable and nuclear expansion, our analysis shows that China's eastern regions will continue to experience an electricity deficit by 2050. This gap could be partly alleviated through hydrogen imports, either to support electricity generation or to meet domestic energy demand; if imports were to cover half of the shortfall, this would represent an additional 275 TWh of hydrogen trade in 2050. The situation in the United States is markedly different: by 2050, no states are projected to face an energy deficit when accounting for renewable resources and planned transmission line expansions. In addition, the country has ambitious nuclear expansion plans and a strategic interconnection with Canada, which together imply that its demand for hydrogen imports will remain minimal.

It is essential to note, however, that these findings rely on the assumption of a fully decarbonised global energy system by 2050, which requires aggressive renewable deployment worldwide. The analysis assumes that countries not identified as importers lack the incentive to export hydrogen. However, in practice, this assumption may not be applicable, as energy security concerns or comparatively lower production costs in neighbouring regions could motivate export activity.

## 4.4 Limitations

This study has limitations when analysing the importing country-specific hydrogen demand and supply potential. It excludes cases involving synthetic fuels trade. Additionally, as the analysis is focused on the import demand, it does not cover hydrogen transport methods, the work by Pieton et al. 2023 provides insights on long-distance transport[113]. Hydrogen demand is considered based on final hydrogen consumption. Consequently, the synthetic fuel market is not considered in this study, which could present a fundamentally different global landscape compared to hydrogen. Further information on synthetic fuel trading and export- and import-oriented countries for PtX/L products addressed in the article by Galimova et al. [17] Furthermore, considerations regarding security of supply were not the focus of this study.

The results may be influenced by the number and type of studies available for each country. Overall, there is a clear preponderance of studies for countries such as Germany, Italy, Japan, and South Korea, reflecting substantial research funding programmes in recent years. This highlights the need for more country-specific analyses in regions where few studies currently exist. To reflect the global hydrogen market in more detail and capture countries' cross-border trade interdependencies for global energy system models, is required. For the in-depth case study analysis, a clear limitation is the lack of data on hydrogen pipeline plans in the United States. However, the country is well-positioned, and intra-state hydrogen exchanges are unlikely to significantly affect the national balance.

Our study focuses on the global renewable hydrogen market, excluding any capacity for other low-carbon hydrogen production pathways. When these are considered, they will further reduce renewable hydrogen import volumes. Other constraints, such as the potential for geological hydrogen storage and existing gas infrastructure, are also not considered and may influence the allocation of renewable hydrogen production in countries that are both exporters and importers. Additionally, advances in hydrogen certification could influence the development and implementation of potential hydrogen trade agreements, which were not the focus of this study.

# 5 Conclusion

The paper examines the potential global hydrogen import in 2050 by identifying the demand and domestic supply of import-oriented countries through a meta-analysis. In total, 16 countries emerged as potential importers. Our findings suggest that global hydrogen trade volumes are unlikely to reach the multi-PWh scale projected by leading consultancies but instead remain in the 576-1,514 TWh range.

The analysis indicates that economies such as Germany, the Netherlands, Belgium and Italy will remain import-oriented, whereas some Asian countries, including South Korea, Indonesia, Vietnam, and the Philippines possess sufficient renewable energy potential to cover their decarbonisation needs. The focus analysis of China shows that some eastern provinces could import hydrogen from other countries due to lower costs, rather than utilising the renewable energy potential from western provinces. The U.S., in contrast, has abundant renewable resources in every state and existing interstate energy infrastructure to meet its domestic demand. Nevertheless, cross-border electricity and hydrogen trade with Canada and Mexico may still offer cost advantages. The intention to have a common clean hydrogen production standard in the region reflects this. Similar regional dynamics could unfold in other large economies, including India, Russia and Mexico, which can be investigated in future research.

Overall, the hydrogen market is likely to remain regional and inter regional rather than global due to the widespread distribution of renewable resources with comparable low electricity production cost, as well as the high transportation costs of hydrogen, both compared to conventional fossil fuel markets. Synthetic fuel markets may evolve towards hydrogen, where production and conversion steps are concentrated in regions with the lowest-cost renewable energy, and where hydrogen is traded globally as an intermediate or final product. This could be due to the high energy-to-mass ratio, the similar chemical characteristics to fossil fuels, and the established trading infrastructure.

Finally, our study highlights important limitations. Some regions remain under-studied, reflecting disparities in data availability and methodological assumptions across existing literature. Global energy system analysis is crucial for capturing interregional interdependencies in future hydrogen trade. Inclusion of other means of hydrogen production from nuclear energy or via SMR with CCS is essential as well and will minimise further the intercontinental hydrogen flows.

*During the preparation of this work, the author(s) used DeepL and ChatGPT to check the wording and improve the readability of the manuscript. After using this tool/service, the author(s) reviewed and edited the content as needed and take(s) full responsibility for the content of the publication.*

Acknowledgment

This work was supported by the HYPAT project, funded by the German Federal Ministry of Education and Research (BMBF). The authors are solely responsible for the content of this publication.

12. Pfennig, M. *et al.* Global GIS-based potential analysis and cost assessment of Power-to-X fuels in 2050. *Applied Energy* **347,** 121289; 10.1016/j.apenergy.2023.121289 (2023).

13. Brändle, G., Schönfisch, M. & Schulte, S. Estimating long-term global supply costs for low-carbon hydrogen. *Applied Energy* **302,** 117481; 10.1016/j.apenergy.2021.117481 (2021).

14. Heuser, P.-M., Grube, T., Heinrichs, H. & Robinius, M. Worldwide Hydrogen Provision Scheme Based on Renewable Energy. *Preprints* (2020).

15. Panchenko, V. A., Daus, Y., Kovalev, A. A., Yudaev, I. V. & Litti, Y. Prospects for the production of green hydrogen: Review of countries with high potential. *International Journal of Hydrogen Energy* **48,** 4551–4571; 10.1016/j.ijhydene.2022.10.084 (2023).

16. Bühler, L. & Scharf, H. Green Hydrogen in the European Union – A large-scale assessment of the supply potential and economic viability. *Applied Energy* **401,** 126587; 10.1016/j.apenergy.2025.126587 (2025).

17. Galimova, T. *et al.* Global trading of renewable electricity-based fuels and chemicals to enhance the energy transition across all sectors towards sustainability. *Renewable and Sustainable Energy Reviews* **183,** 113420; 10.1016/j.rser.2023.113420 (2023).

18. Heuser, P.-M., Grube, T., Heinrichs,, H., Robinius, M. & Stolten, D. Worldwide Hydrogen Provision Scheme Based on Renewable Energy. *Preprints* (2020).

19. IEE. Power-to-X Potenzialatlas. Available at https://maps.iee.fraunhofer.de/ptx-atlas/ (2021).

20. Heuser, P.-M., Ryberg, D. S., Grube, T., Robinius, M. & Stolten, D. Techno-economic analysis of a potential energy trading link between Patagonia and Japan based on CO2 free hydrogen. *International Journal of Hydrogen Energy* **44,** 12733–12747; 10.1016/j.ijhydene.2018.12.156 (2019).

21. IEA. Korea 2020 Energy Policy Review, 2020.

22. Newborough, M. & Cooley, G. Developments in the global hydrogen market: Electrolyser deployment rationale and renewable hydrogen strategies and policies. *Fuel Cells Bulletin* **2020,** 16–22; 10.1016/S1464-2859(20)30486-7 (2020).